\documentclass[12pt]{article}
\usepackage{epsfig,amsmath,amssymb,stmaryrd,enumerate}
\usepackage{graphicx}

\advance \textheight by \topskip
\advance \textheight by 1pt
\makeatother

\begin{document}

\title{Difference Krichever--Novikov operators}
\author{{Gulnara S.~Mauleshova and Andrey E.~Mironov}}
\date{}
\maketitle

\begin{abstract}
In this paper we study commuting difference operators of rank two. We introduce an equation on potentials $V(n),W(n)$ of the
difference operator $L_4=(T+V(n)T^{-1})^2+W(n)$ and some additional data. With the help of this equation we find the first examples
of commuting difference operators of rank two corresponding to spectral curves of higher genus.
\end{abstract}

\section{Introduction and main results}

 I.M.~Krichever and S.P.~Novikov \cite{KN1}, \cite{KN2} discovered a remarkable class of solutions of soliton
 equations --- algebro--geometric solutions of rank $l>1$.
 This class is determined by the following condition: common eigenfunctions of auxiliary commuting ordinary differential or difference operators form a
 vector bundle of rank $l$ over the spectral curve. Rank two solutions of the Kadomtsev--Petviashvili (KP) equation and 2D-Toda chain corresponding
 to the spectral curves of genus $g=1$ were found in \cite{KN1}, \cite{KN2}. To find higher rank solutions one has to find higher rank commuting
 operators and their appropriate deformations. The problem of classification of commuting differential and difference operators
 was solved in \cite{KN2}--\cite{K1};
however, finding the operators themselves has remained an open problem.
Moreover, no examples of commuting differential operators of rank $l>1$ at $g>1$ were known before the recent paper \cite{Mir} (see also \cite{Mir3}).

In this paper we study commuting difference operators.
We denote by $L_k,L_s$ the operators of orders $k=N_-+N_+$ and $s=M_-+M_+$
$$
 L_k=\sum^{N_+}_{j=N_-} u_j(n)T^j, \quad
 L_s=\sum^{M_+}_{j=M_-}v_j(n)T^j, \quad
 n \in \mathbb Z,
$$
where $T$ is the shift operator. The condition of their commutativity is equivalent to a complicated system of nonlinear
difference equations on the coefficients. These equations have been studied since the beginning of the 20th century (see \cite{W}). An analogue of
the Burchnall--Chaundy lemma \cite{BC} holds. Namely, if $L_kL_s=L_sL_k$, then there exists a nonzero polynomial
$F(z,w)$ such that $F(L_k,L_s)=0$
\cite{K3}. The polynomial $F$ defines the {\it spectral curve}
$$
 \Gamma=\{(z,w)\in {\mathbb C}^2| F(z,w)=0\}.
$$
The spectral curve parametrizes common eigenvalues, i.e. if
$$
 L_k\psi=z\psi, \qquad L_s\psi=w\psi,
$$
then $(z,w)\in \Gamma.$  The dimension of the space of common eigenfunctions with the fixed eigenvalues is called the {\it rank} of the pair $L_k,L_s$
$$
 l={\rm dim}\{\psi:L_k\psi=z\psi, \ \ L_s\psi=w\psi\},
$$
where the point $(z,w)\in \Gamma$ is in general position. Thus the
spectral curve and the rank are defined exactly the same way as
in the case of the differential operators.

Before discussing difference operators we briefly discuss differential operators. The first important results relating to commuting
differential operators of rank $l>1$ were obtained by J. Dixmier
\cite{D} and V.G. Drinfel'd \cite{Dr}.  The operators of
rank $l$ with periodic coefficients were studied in \cite{Nov}. Rank two operators at $g=1$ were found by I.M. Krichever and S.P. Novikov
\cite{KN1}. These operators were studied in
\cite{BE}--\cite{PW} (see also \cite{M1}--\cite{Z} for $g=2$--$4, l=2,3$). At  $g=1, l=3$ operators were
found in \cite{Mokh}. Methods of \cite{Mir} allow to construct and study
higher rank operators at $g>1$
\cite{Dav}--\cite{Mokh1}.

The maximal commutative ring of difference
operators containing $L_k$ and $L_s$ is isomorphic to the ring of
meromorphic functions on an algebraic spectral curve $\Gamma$ with poles
$q_1,\ldots,q_m\in\Gamma$ (see \cite{KN2}).
Such operators are called {\it $m$-points operators}. We note that any ring of
commuting differential operators is isomorphic to a ring of
meromorphic functions on a spectral curve with a unique pole.
The commuting difference
operators of rank one were found by I.M.~Krichever \cite{K3} and D.
Mumford \cite{Mum}. Eigenfunctions (Baker--Akhiezer functions) and coefficients of such operators
can be found explicitly with the help of
theta-functions of the Jacobi varieties of spectral curves. In the case of $l>1$ eigenfunctions cannot be
found explicitly. Finding such operators is still an open problem.
Rank two one-point operators at $g=1$ were found in \cite{KN2}, operators with
polynomial coefficients among them were obtained in \cite{Mir1}.

In this paper we consider one-point operators of rank two $L_4, L_{4g+2}$
corresponding to the hyperelliptic spectral curve $\Gamma$
\begin{equation}
\label{eq1}
w^2=F_g(z)=z^{2g+1}+c_{2g}z^{2g}+c_{2g-1}z^{2g-1}+...+c_0,
\end{equation}
herewith
\begin{equation}
\label{eq2}
L_4=\sum^{2}_{i=-2}
u_i(n)T^i, \qquad L_{4g+2}=\sum^{2g+1}_{i=-(2g+1)}v_i(n)T^i,\qquad u_2=v_{2g+1}=1,
\end{equation}
\begin{equation}
\label{eq3}
L_4\psi=z\psi, \qquad L_{4g+2}\psi=w\psi,\qquad \psi=\psi(n,P),\qquad P=(z,w)\in\Gamma.
\end{equation}
Common eigenfunctions of $L_4$ and $L_{4g+2}$ satisfy the equation
\begin{equation}
\label{eq4}
\psi(n+1,P)=\chi_1(n,P)\psi(n-1,P)+\chi_2(n,P)\psi(n,P),
\end{equation}
where $\chi_1(n,P)$ and $\chi_2(n,P)$ are rational functions on $\Gamma$ having
$2g$ simple poles, depending on $n$ (see \cite{KN2}). The function $\chi_2(n,P)$
additionally has a simple pole at $q=\infty$.
To find $L_4$ and $L_{4g+2}$ it is sufficient to find $\chi_1$ and $\chi_2.$ Let $\sigma$ be the holomorphic involution on $\Gamma, \ \ \sigma(z,w)=\sigma(z,-w).$
The main results of this paper are Theorems 1--4.


${\ }$

\noindent\textbf{Theorem 1}
{\it If
\begin{equation}
\label{eq5}\chi_1(n,P)=\chi_1(n,\sigma(P)),\qquad
\chi_2(n,P)=-\chi_2(n,\sigma(P)),
\end{equation}
then $L_4$ has the form
\begin{equation}
\label{eq6}L_4=(T+V_nT^{-1})^2+W_n,
\end{equation}
where
\begin{equation}
\label{eq7}\chi_1=-V_n\frac{Q_{n+1}}{Q_{n}},\qquad
\chi_2=\frac{w}{Q_n},\qquad Q_n(z)=z^g+\alpha_{g-1}(n)z^{g-1}+\ldots+\alpha_0(n).
\end{equation}
Functions $V_n, W_n, Q_n$ satisfy
\begin{eqnarray}
\label{eq8}
F_g(z)=Q_{n-1}Q_{n+1}V_n+Q_{n}Q_{n+2}V_{n+1}+Q_nQ_{n+1}(z-V_n-V_{n+1}-W_n).
\end{eqnarray}}
${\ }$

\noindent In Theorem 1 and further we use the notations  $V_n=V(n),W_n=W(n)$. It is a remarkable fact that
(\ref{eq8}) can be linearized. Namely, if we replace $n\rightarrow n+1$ and take the difference with (\ref{eq8}),
then the result can be divided by
$Q_{n+1}(z).$ Finally we obtain the linear equation on $Q_n(z)$.

${\ }$

\noindent\textbf{Corollary 1} {\it
Functions $Q_n(z), V_n, W_n$ satisfy
\begin{equation}
\label{eq9}
 Q_{n-1}V_n+Q_n(z-V_n-V_{n+1}-W_{n})-Q_{n+2}(z-V_{n+1}-V_{n+2}-W_{n+1})-Q_{n+3}V_{n+2}=0.
\end{equation}}
${\ }$

\noindent At $g=1$, the equation (\ref{eq8}) allows us to express $V_n, W_n$ via a functional parameter $\gamma_n.$

${\ }$

\noindent\textbf{Corollary 2} {\it
The operator
$
 L_4=(T+V_nT^{-1})^2+W_n,
$ where
\begin{equation}
\label{eq10}
V_n=\frac{F_1(\gamma_n)}{(\gamma_n-\gamma_{n-1})(\gamma_n-\gamma_{n+1})},\qquad
W_n=-c_2-\gamma_n-\gamma_{n+1},
\end{equation}
commutes with
$$
L_6=T^3+(V_n+V_{n+1}+V_{n+2}+W_{n}-\gamma_{n+2})T+
$$
$$
+V_n(V_{n-1}+V_n+V_{n+1}+W_n-\gamma_{n-1})T^{-1}+V_{n-2}V_{n-1}V_nT^{-3}.
$$
The spectral curve of $L_4, L_6$ is $w^2=F_1(z).$}

${\ }$

In the theory of commuting ordinary differential operators there are equations which are similar to (\ref{eq8}), (\ref{eq9}). Let us compare (\ref{eq8}),
(\ref{eq9}) with their smooth analogues.
First, we consider the one-dimensional finite-gap Schr\"odinger operator ${\cal L}_2=-\partial_x^2+{\cal V}(x)$ commuting with a differential operator
${\cal L}_{2g+1}$ of order $2g+1$. The theory of such operators is closely related to the theory of periodic and
quasiperiodic solutions of the Korteweg--de Vries equation (see \cite{N}--\cite{IM}). Denote by $\psi$ a common eigenfunction
$$
 (-\partial_x^2+{\cal V}(x))\psi=z\psi,\qquad {\cal L}_{2g+1}\psi=w\psi.
$$
The point $P=(z,w)$ belongs to the spectral curve (\ref{eq1}). Function $\psi(x,P)$ satisfies
$$
 \psi'(x,P)=i{\cal \chi}_0(x,P)\psi(x,P),
$$
where
$$
 \chi_0=\frac{{\cal Q}_x}{2i{\cal Q}}+\frac{w}{{\cal Q}},\qquad {\cal Q}=z^g+\alpha_{g-1}(x)z^{g-1}+\ldots+\alpha_0(x).
$$
Polynomial ${\cal Q}$ satisfies the equation
$$
 4F_g(z)=4(z-{\cal V}){\cal Q}^2-({\cal Q}_x)^2+2{\cal Q}{\cal Q}_{xx},
$$
which is linearized as well as (\ref{eq8}) (see \cite{DMN}, \cite{GD})
$$
 {\cal Q}_{xxx}-4{\cal Q}_x({\cal V}-z)-2{\cal V}_x{\cal Q}=0.
$$
Equations (\ref{eq8}), (\ref{eq9}) are analogues of the last two.

Let us consider one more example. We denote by ${\cal L}_4$, ${\cal L}_{4g+2}$ rank two commuting differential operators with the spectral curve (\ref{eq1}).
The common eigenfunctions of ${\cal L}_4$ and
${\cal L}_{4g+2}$ satisfy
$$
\psi^{''}=\chi_1(x,P)\psi'+\chi_0(x,P)\psi.
$$
In \cite{Mir} it was proved that  ${\cal L}_4$ is self-adjoint if and only if
$\chi_1(x,P)=\chi_1(x,\sigma (P)),$
herewith
$$
 {\cal L}_4=(\partial_x^2+{\cal V}(x))^2+{\cal W}(x),
$$
$$
 \chi_0=-\frac{1}{2}\frac{{\cal Q}_{xx}}{{\cal Q}}+\frac{w}{{\cal Q}}-{\cal V}, \qquad \chi_1=\frac{{\cal Q}_x}{{\cal Q}},
 \qquad {\cal Q}=z^g+\alpha_{g-1}(x)z^{g-1}+\ldots+\alpha_0(x),
$$
polynomial ${\cal Q}$
satisfies
\begin{equation}
 \label{eq11}
4F_g(z)=4(z-{\cal W}){\cal Q}^2-4{\cal V}({\cal Q}_x)^2+{\cal Q}_{xx}^2-2{\cal Q}_x{\cal Q}_{xxx}+2{\cal Q}(2{\cal V}_x{\cal Q}_x+4{\cal V}{\cal Q}_{xx}+
{\cal Q}_{xxxx}),
\end{equation}
and also satisfies
\begin{equation}
 \label{eq12}
 \partial_x^5{\cal Q}+4{\cal V}{\cal Q}_{xxx}+2{\cal Q}_x(2z-2{\cal W}-{\cal V}_{xx})+6{\cal V}_x{\cal Q}_{xx}-2{\cal Q}{\cal W}_x=0.
\end{equation}
Equations (\ref{eq8}), (\ref{eq9}) are discrete analogues of (\ref{eq11}), (\ref{eq12}).

Theorem 1 allows us to construct the examples.

${\ }$

\noindent\textbf{Theorem 2} {\it
The operator
$$
L^{^\sharp}_4=(T+(r_3n^3+r_2n^2+r_1n+r_0)T^{-1})^2+g(g+1)r_3n, \qquad r_3\ne 0
$$
commutes with a difference operator $L^{^\sharp}_{4g+2}.$}

${\ }$

\noindent\textbf{Theorem 3} {\it
The operator
$$
 L^{^\checkmark}_4=(T+(r_1a^n+r_0)T^{-1})^2+r_1(a^{2g+1}-a^{g+1}-a^g+1)a^{n-g},\qquad r_1, a\ne 0,
$$
where $a^{2g+1}-a^{g+1}-a^g+1\ne 0$,
commutes with a difference operator $L^{^\checkmark}_{4g+2}.$}

${\ }$

\noindent\textbf{Theorem 4} {\it
The operator
$$
L^{^\natural}_4=(T+(r_1\cos(n)+r_0)T^{-1})^2-4r_1\sin(\frac{g}{2})\sin(\frac{g+1}{2})\cos(n+\frac{1}{2}),\qquad r_1\ne 0
$$
commutes with a difference operator $L^{^\natural}_{4g+2}.$}

${\ }$

\noindent In Section 2 we recall the Krichever--Novikov equations on Tyurin parameters.

\noindent In Section 3 we prove Theorems 1--4 and consider examples.

\noindent In Appendix we consider the differential--difference system on $V_n(t), \ W_n(t)$
\begin{equation}
\label{eq26}
\dot{V}_n=V_n(W_{n-1}-W_n+V_{n-1}-V_{n+1}),
\end{equation}
\begin{equation}
\label{eq27}
\dot{W}_n=(W_{n}-W_{n-1})V_n+(W_{n+1}-W_n)V_{n+1}.
\end{equation}
From (\ref{eq26}), (\ref{eq27}) it follows that $\varphi_n(t)$, where $e^{\varphi_n(t)}=V_n(t)$, satisfies the generalized Toda chain
$$
 \ddot{\varphi}_n=e^{\varphi_{n-2}+\varphi_{n-1}}-e^{\varphi_{n-1}+\varphi_{n}}+e^{\varphi_{n+1}+\varphi_{n+2}}-e^{\varphi_{n+1}+\varphi_{n}}.
$$
From (\ref{eq26}), (\ref{eq27}) it follows also that
$$
 [L_4,\partial_t-V_{n-1}(t)V_n(t)T^{-2}]=0,
$$
where $L_4=(T+V_n(t)T^{-1})^2+W_n(t).$
Following \cite{KN1}, \cite{KN2} we call the solution $V_n(t),W_n(t)$ of (\ref{eq26}), (\ref{eq27}) {\it the solution of rank two}, if additionally
$[L_4,L_{4g+2}]=0$ for some difference operator $L_{4g+2}$.
In the case of rank two solutions an evolution equation on $Q_n(t)$ is obtained in Theorem 5. At $g=1$ this equation is reduced to a discrete
analogue of the Krichever--Novikov equation, which appeared in the theory of rank two solutions of KP.

${\ }$

\section{Discrete dynamics of the Tyurin parameters}

As mentioned above, in the case of rank one operators the eigenfunctions can be found explicitly in terms of
theta-functions of the Jacobi varieties of spectral curves. Let us consider the simplest example.
Let $\Gamma$ be an elliptic curve $\Gamma={\mathbb C}/\{{\mathbb Z}+\tau{\mathbb Z}\}, \tau\in{\mathbb C}, {\rm Im}\tau>0,$ and
$\theta(z)$ the theta-function $\theta(z)=\sum_{n\in{\mathbb Z}}\exp(\pi i n^2\tau+2\pi i n z).$
The Baker--Akhiezer function has the form
\begin{equation}
\label{eq14}
\psi(n,z)=\frac{\theta(z+c+nh)}{\theta(z)}\left(\frac{\theta(z-h)}{\theta(z)}\right)^n,\qquad c,h\notin \{{\mathbb Z}+\tau{\mathbb Z}\}.
\end{equation}
For the meromorphic function $\lambda=\frac{\theta(z-a_1)\ldots(z-a_k)}{\theta^k(z)},$ $a_1+\ldots+a_k=0$ there is a unique operator
$
L(\lambda)=v_k(n)T^k+\ldots+v_0(n)
$
such that
$
L(\lambda)\psi=\lambda\psi.
$
Coefficients of $L(\lambda)$ can be found from the last identity (see \cite{MN}). Operators $L(\lambda)$ for different $\lambda$ form a commutative
ring of difference operators.
 This example can be generalized from the elliptic spectral curves to the principle polarized abelian spectral varieties. It allows to
construct commuting difference operators in several discrete variables with matrix coefficients (see \cite{MN}).

At $l>1$ common eigenfunctions cannot be found explicitly.
This is the main difficulty for constructing higher rank operators and higher rank solutions of the 2D-Toda chain.
Recall the needed results of \cite{KN2}.
One-point commuting operators of rank $l$ have the form
$$
L=\sum^{Nr_+}_{i=-Nr_-}u_i(n)T^i, \qquad
A=\sum^{Mr_+}_{i=-Mr_-}v_i(n)T^i,
$$
where $l=r_{-}+r_{+},  (N,M)=1$.
Consider the space ${\cal H}(z)$ of solutions of the equation
$Ly=zy.$ We have ${\rm dim} {\cal H}(z)=N(r_{-}+r_{+}).$
The operator $A$ defines
the linear operator $A(z)$ on ${\cal H}(z).$
Let us choose the basis $\varphi^i(n)$ in ${\cal H}(z)$,
satisfying the normalization conditions
$
 \varphi^i(n)=\delta_{in},  -Nr_{-}\leq i,n<Nr_{+}.
$
The components of $A(z)$ in the basis $\varphi^i(n)$ are polynomials in $z$.
The characteristic polynomial of
$A(z)$ has the form
$
{\rm det}(w-A(z))=R^l(w,z).
$
Polynomial $R$ defines the spectral curve $\Gamma$, i.e.
$$
 L\psi=z\psi, \qquad A\psi=w\psi,\qquad R(z,w)=0.
$$
Common eigenfunctions of $L$ and $A$ form a vector bundle of rank
$l$ over the affine part of $\Gamma$. Let us choose the basis in the space of common eigenfunctions such that
$$
\psi^i_n(P)=\delta_{i,n}, \qquad -r_{-}\leq i,n<r_+, \qquad P=(z,w)\in\Gamma.
$$
Functions $\psi^i_n(P)$ have the pole divisor $\gamma=\gamma_1+\ldots+\gamma_{lg}$ of degree $lg$.
We have the following identities
$$
 \alpha_s^j{\rm Res}_{\gamma_s}\psi^i_n(P)=\alpha_s^i{\rm Res}_{\gamma_s}\psi^j_n(P).
$$
The pair $(\gamma,\alpha)$ is called the {\it Tyurin parameters}, where
$\alpha$ is the set of vectors
$$
 \alpha_1,\dots,\alpha_{lg},\qquad \alpha_s=(\alpha_s^{-r_-},\dots,\alpha_s^{r_+-1}).
$$
The Tyurin parameters define a stable holomorphic vector bundle on $\Gamma$ of degree $lg$ with holomorphic sections $\zeta_{-r_-},\dots,\zeta_{r_+-1}$, where
$\gamma$ is the divisor of their linear dependence
$
 \sum_{j=-r_-}^{r_+-1}\alpha_s^j\zeta_j(\gamma_s)=0.
$
Let $\Psi(n,P)$ be the Wronski matrix with the components
$
\Psi^{ij}(n,P)=\psi_{n+j}^{i}(P),
-r_{-}\leq i,j<r_+.
$
Function ${\rm det}\Psi(n,P)$ is holomorphic in the neighbourhood of $q=\infty$. The pole divisor of ${\rm det}\Psi(n,P)$ is $\gamma$, the zero divisor
of ${\rm det}\Psi(n,P)$ is $\gamma(n)=\gamma_1(n)+\dots+\gamma_{lg}(n),$
herewith $\gamma(0)=\gamma$.
Consider the matrix function
$\chi(n,P)=\Psi(n+1,P)\Psi^{-1}(n,P),$
$$
\chi(n,P)=\left(%
\begin{array}{ccccc}
  0 & 1 & 0 & \ldots & 0 \\
  0 & 0 & 1 & \ldots & 0 \\
 \cdots&\cdots&\cdots&\cdots&\cdots \\
  0 & 0 & 0 & \ldots & 1 \\
  \chi_{-r_-}(n,P) & \chi_{-r_{-}+1}(n,P) & \chi_{-r_{-}+2}(n,P) & \ldots & \chi_{r_{+}-1}(n,P) \\
\end{array}
\right).
$$
In the neighbourhood of $q$ we have
$
 \chi_i(n,k)=k^{-1}\delta_{i,0}-f_i(n,k),
$
where $k$ is a local parameter near $q$, $f_i(n,k)$ is an analytical function in the neighbourhood of $q$.

 ${\ }$

\noindent\textbf{Theorem} (I.M. Krichever, S.P. Novikov)

\noindent{\it
The matrix function $\chi(n,P)$ has simple poles in $\gamma_s(n)$, and
\begin{equation}
\label{eq15}
\alpha_s^j(n){\rm Res}_{\gamma_s(n)}\chi_i(n,P)=\alpha_s^i(n){\rm Res}_{\gamma_s(n)}\chi_j(n,P).
\end{equation}
Points $\gamma_s(n+1)$ are zeros of ${\rm det}\chi(n,P)$
\begin{equation}
\label{eq16} {\rm det}\chi(n,\gamma_s(n+1))=0.
\end{equation}
Vectors $\alpha_j(n+1)=(\alpha_s^{-r_-}(n+1),\dots,\alpha_s^{r_+-1}(n+1))$ satisfy
\begin{equation}
\label{eq17} \alpha_s(n+1)\chi(n,\gamma_s(n+1))=0.
\end{equation}}

Equations (\ref{eq15})--(\ref{eq17}) define the discrete dynamics of the Tyurin parameters.
In \cite{KN2} solutions of (\ref{eq15})--(\ref{eq17}) are found at $g=1, l=2$. The corresponding operators in
the simplest case have the form
$$
 L=L_2^2-\wp(\gamma_n)-\wp(\gamma_{n-1}),
$$
$L_2$ is the difference Schr\"odinger operator $L_2=T+v_n+c_nT^{-1}$
with the coefficients
$$
 c_n=\frac{1}{4}(s_{n-1}^2-1)F(\gamma_n,\gamma_{n-1})F(\gamma_{n-2},\gamma_{n-1}),\qquad
 v_n=\frac{1}{2}(s_{n-1}F(\gamma_n,\gamma_{n-1})-s_nF(\gamma_{n-1},\gamma_n)),
$$
where
$$
 F(u,v)=\zeta(u+v)-\zeta(u-v)-2\zeta(v).
$$
Here, $\wp(u),$ $\zeta(u)$ are the Weierstrass functions, $s_n,$
$\gamma_n$ are the functional parameters.

${\ }$

\section{Proof of Theorems 1--4}

Let $\Gamma$ be the hyperelliptic spectral curve (\ref{eq1}), $L_4,$ $L_{4g+2}$ are operators of the form (\ref{eq2}) with the properties (\ref{eq3}).
Matrix $\chi(n,P)=\Psi(n+1,P)\Psi^{-1}(n,P)$ has the form
$$
 \chi(n,P)=\left(
\begin{array}{ccccc}
  0 & 1 \\
  \chi_1(n,P) & \chi_2(n,P)\\
\end{array}
\right).
$$
The functions $\chi_1,$ $\chi_2$ have the following expansions in the neighbourhood of $q=\infty$:
\begin{equation}
\label{eq18}
\chi_1(n)=b_0(n)+b_1(n)k+\ldots, \qquad
\chi_2(n)=1/k+e_0(n)+e_1(n)k+\ldots,
\end{equation}
where $k=\frac{1}{\sqrt{z}}.$

${\ }$

\noindent\textbf{Lemma 1} {\it
The operator
$$
L_4=T^2+u_1(n)T+u_0(n)+u_{-1}(n)T^{-1}+u_{-2}(n)T^{-2}
$$
has the coefficients:
$$
u_1(n)=-e_0(n)-e_0(n+1),
\quad
u_0(n)=e_0^2(n)+e_1(n)-e_1(n+1)-b_0(n)-b_0(n+1),
$$
$$
u_{-1}(n)=b_0(n)\left(e_0(n)+e_0(n-1)-\frac{b_1(n-1)}{b_0(n-1)}\right)-b_1(n),\quad
u_{-2}(n)=b_0(n)b_0(n-1).
$$
If $b_1(n)=0$, $e_0(n)=0$, then $L_4$ can be written in the form (\ref{eq6}), where
$$
V_n=-b_0(n),\qquad W_n=-e_1(n)-e_1(n+1).
$$}
${\ }$

\noindent\textbf{Proof} Using (\ref{eq4}) let us express
$\psi_{n+2}(P)$ and $\psi_{n-2}(P)$ via $\psi_{n-1}(P),$
$\psi_n(P),$ $\chi_1(n,P),$ $\chi_2(n,P)$
$$
\psi_{n+2}=\psi_{n-1}\chi_1(n)\chi_2(n+1)+\psi_n(\chi_1(n+1)+\chi_2(n)\chi_2(n+1)),\quad
 \psi_{n-2}=\frac{\psi_n-\psi_{n-1}\chi_2(n-1)}{\chi_1(n-1)},
$$
and substitute it in
$L_4\psi_n=z\psi_n$. We get
$
P_1(n,P)\psi_n(P)+P_2(n,P)\psi_{n-1}(P)=z\psi_n(P),
$
where
$$
P_1(n)=\chi_1(n+1)+\chi_2(n+1)\chi_2(n)+u_1(n)\chi_2(n)+u_0(n)+\frac{u_{-2}(n)}{\chi_1(n-1)},
$$
$$
P_2(n)=\chi_2(n+1)\chi_1(n)+u_1(n)\chi_1(n)+u_{-1}(n)-u_{-2}(n)\frac{\chi_2(n-1)}{\chi_1(n-1)}.
$$
Consequently we have
\begin{equation}
\label{eq19}
P_1=z=\frac{1}{k^2}, \qquad P_2=0.
\end{equation}
From (\ref{eq18}), (\ref{eq19}) it follows that
$$
 P_1-\frac{1}{k^2}=\frac{e_0(n)+e_0(n+1)+u_1(n)}{k}+\left(b_0(n+1)+e_0(n)e_0(n+1)+e_1(n)+\right.
$$
$$
 \left.e_1(n+1)+u_0(n)+e_0(n)u_1(n)+\frac{u_{-2}(n)}{b_0(n-1)}\right)+O(k)=0,
$$
$$
 P_2=\frac{b_0(n)-\frac{u_{-2}(n)}{b_0(n-1)}}{k}+\left(b_1(n)+b_0(n)e_0(n+1)+b_0(n)u_1(n)+\right.
$$
$$
 \left.u_{-1}(n)+\frac{b_1(n-1)u_{-2}(n)}{b_0^2(n-1)}-\frac{e_0(n-1)u_{-2}(n)}{b_0(n-1)}\right)+O(k)=0.
$$
This yields the formulas for the coefficients of $L_4.$ By direct calculations one can check that if  $b_1(n)=e_0(n)=0$, then $L_4$ has the form (\ref{eq6}).
Lemma 1 is proved.

${\ }$

Thus if $\chi_1,$ $\chi_2$ satisfy (\ref{eq5}),
then $b_1(n)=e_0(n)=0,$ and hence $L_4$ has the form (\ref{eq6}).
Operators $L_4-z$ and $L_{4g+2}-w$ have the common right divisor $T-\chi_2(n)-\chi_1(n)T^{-1},$ i.e.
$$
L_4-z=l_1(T-\chi_2(n)-\chi_1(n)T^{-1}), \qquad
L_{4g+2}-z=l_2(T-\chi_2(n)-\chi_1(n)T^{-1}),
$$
where $l_1$ and $l_2$ are operators of orders $2$ and $4g$.
Let us assume that (\ref{eq5}) holds. Then
$$
(T+V_nT^{-1})^2+W_n-z=(T+\chi_2(n+1)-\frac{V_{n-1}V_n}{\chi_1(n-1)}T^{-1})(T-\chi_2(n)-\chi_1(n)T^{-1}),
$$
where $\chi_1$, $\chi_2$ satisfy the equations
\begin{eqnarray}
\label{eq20}
 V_{n-1}V_n+\chi_1(n-1)(V_n+V_{n+1}-z+
 W_n+\chi_1(n+1)+\chi_2(n)\chi_2(n+1))=0,
\end{eqnarray}
\begin{equation}
\label{eq21}
 -V_{n-1}V_n\chi_2(n-1)+\chi_1(n-1)\chi_1(n)\chi_2(n+1)=0.
\end{equation}
We have
$
 {\rm det}\chi(n,P)=-\chi_1(n,P)={\rm det}\Psi(n+1,P)({\rm det}\Psi(n,P))^{-1}.
$
The degree of the zero divisor $\gamma(n)$ of ${\rm det}\Psi(n,P)$ is $2g.$ Since $\chi_1$ is invariant under the involution $\sigma$,
the divisor $\gamma(n)$ has the form
$$
\gamma(n)=\gamma_1(n)+\sigma\gamma_1(n)+\ldots+\gamma_g(n)+\sigma\gamma_g(n).
$$
Let $\gamma_i(n)$ have the coordinates
$(\mu_i(n),w_i(n)).$
We introduce the polynomial in $z$
$$
 Q_n=(z-\mu_1(n))\ldots(z-\mu_g(n)).
$$
From (\ref{eq16}) we have
$
 \chi_1(n,P)=b_0(n)\frac{Q_{n+1}}{Q_n},
$
where $b_0(n)$ is some function. In the neighbourhood of $q$ we have
$$
\chi_1=b_0(n)+b_2(n)k^2+O(k^4).
$$
By Lemma 1 $V_n=-b_0(n),$ so we get
$
\chi_1(n,P)=-V_n\frac{Q_{n+1}}{Q_n}.
$

Since the pole divisor of $\chi_2(n,P)$ is $\gamma(n)$ and in the neighborhood of $q$ we have (\ref{eq18}), then
$
 \chi_2(n,P)=\frac{w}{Q_n}.
$

If $\chi_1(n,P)=-V_n\frac{Q_{n+1}}{Q_n}$ and $\chi_2(n,P)=\frac{w}{Q_n}$, then (\ref{eq21}) holds identically, and (\ref{eq20}) is reduced
to (\ref{eq8}). Theorem 1 is proved.
${\ }$

To prove Theorems 2--4 it is sufficient to prove that for potentials $V_n, W_n$ from Theorems 2--4 there are polynomials $Q_n(z)$ of degree $g$ in
$z$ which satisfy (\ref{eq9}) (and hence satisfy (\ref{eq8})).

\subsection{Theorem 2}
Let
$V_n=r_3n^3+r_2n^2+r_1n+r_0,$ $W_n=g(g+1)r_3n,$
then (\ref{eq9}) takes the form
$$
 Q_{n-1}(n^3r_3+n^2r_2+nr_1+r_0)+
 Q_n(z-2n^3r_3-n^2(2r_2+3r_3)-n(2r_1+2r_2+3r_3+g(g+1)r_3)
$$
$$
 -\left.(2r_0+r_1+r_2+r_3)\right)-Q_{n+2}\left(z-2n^3r_3-n^2(2r_2+9r_3)\right.-
$$
$$
 \left.n(2r_1+6r_2+15r_3+g(1+g)r_3)-(2r_0+3r_1+5r_2+9r_3+g(g+1)r_3)\right)-
$$
\begin{equation}
\label{eq22}
 Q_{n+3}\left(n^3r_3+n^2(r_2+6r_3)+n(r_1+4r_2+12r_3)+r_0+2r_1+4r_2+8r_3\right)=0.
\end{equation}
Let us take the following ansatz for $Q_n(z)$
$$
 Q_n=\delta_gn^g+\ldots+\delta_1n+\delta_0, \qquad
\delta_i=\delta_i(z),
$$
then (\ref{eq22}) can be rewritten in the form
$$
 \beta_{g+3}(z)n^{g+3}+\beta_{g+2}(z)n^{g+2}+\ldots+\beta_0(z)=0
$$
for some $\beta_s(z)$. Potentials $V_n,$ $W_n$ have the following remarkable properties: it turns out that
$$
\beta_{g}=\beta_{g+1}=\beta_{g+2}=\beta_{g+3}=0
$$
automatically (this can be checked by direct calculations). From (\ref{eq22}) we find $\beta_s$
$$
 \beta_s=r_3(2s+1)(g(g+1)-s(s+1))\delta_s+\sum_{m=1}^{g}\left((-1)^m\left({\rm C}_{s+m}^mr_0-{\rm C}_{s+m}^{m+1}r_1\right.\right.+
$$
$$
 \left.{\rm C}_{s+m}^{m+2}r_2-{\rm C}_{s+m}^{m+3}r_3\right)+2^m\left({\rm C}_{s+m}^m(2r_0+3r_1+5r_2+9r_3+g(g+1)r_3-z)+\right.
$$
$$
\left.2{\rm C}_{s+m}^{m+1}(2r_1+6r_2+15r_3+g(g+1)r_3)+4{\rm C}_{s+m}^{m+2}(2r_2+9r_3)+16{\rm C}_{s+m}^{m+3}r_3\right)-
$$
$$
 3^m\left({\rm C}_{s+m}^m(r_0+2r_1+4r_2+8r_3)+3{\rm C}_{s+m}^{m+1}(r_1+4r_2+12r_3)+\right.
$$
$$
 3^m\left({\rm C}_{s+m}^m(r_0+2r_1+4r_2+8r_3)+3{\rm C}_{s+m}^{m+1}(r_1+4r_2+12r_3)+\right.
$$
\begin{equation}
\label{eq23}
\left.\left.9{\rm C}_{s+m}^{m+2}(r_2+6r_3)+27{\rm C}_{s+m}^{m+3}r_3\right)\right)\delta_{s+m},
\end{equation}
where $0\leq s<g-1, \ \ {\rm C}_m^k=\frac{m!}{k!(m-k)!}$ at $m\geq k,$ ${\rm C}_m^k=0$ at $m<k,$ $\delta_g$ is a constant and $\delta_s=0,$ if
$s>g.$ From $\beta_s=0$ we express $\delta_s$ via $\delta_{s+1},\ldots, \delta_g.$ In particular,
$$
 \delta_{g-1}=\frac{\delta_g(2g^2r_2+g(g+1)r_3+2z)}{2(2g-1)r_3}.
$$
For a suitable $\delta_g$ we have $Q_n=z^g+\alpha_{g-1}(n)z^{g-1}+\ldots+\alpha_0(n).$
So we proved that there exists $Q_n$ satisfying (\ref{eq9}). Theorem 2 is proved.

In \cite{Mir} it was proved that
$$
 {\cal L}_4^{^\sharp}=(\partial_x^2+r_3x^3+r_2x^2+r_1x+r_0)^2+g(g+1)r_3x
$$
commutes with a differential operator ${\cal L}_{4g+2}^{^\sharp}$ of order $4g+2$. The operator
$L^{^\sharp}_4$ is a discrete analogue of ${\cal L}_4^{^\sharp}.$ At $g=1$ the operators ${\cal L}_4^{^\sharp},$ ${\cal L}_6^{^\sharp}$
were found by Dixmier \cite{D}. The operators ${\cal L}_4^{^\sharp},$ ${\cal L}_{4g+2}^{^\sharp}$ define a commutative subalgebra in the first Weyl algebra.

Let us consider the algebra $W$ generated by two elements $p$ and $q$ with the relation $[p,q]=p.$
Since $[T,n]=T$, the algebra is isomorphic to the algebra of difference operators with polynomial coefficients. The algebra $W$ has the following automorphisms
$$
 H: W \rightarrow W, \qquad H(p)=p, \qquad H(q)=q+G(p),
$$
where $G$ is an arbitrary polynomial. Operators $L^{^\sharp}_4$, $L^{^\sharp}_{4g+2}$ define the commutative subalgebra in $W$. Consequently,
if we replace  $n\rightarrow n+G(T)$ in $L^{^\sharp}_4$, $L^{^\sharp}_{4g+2}$, then we obtain the new commuting difference operators with polynomial coefficients.


\subsection{Theorem 3}
Let $V_n=r_1a^n+r_0$, $W_n=(a^{2g+1}-a^{g+1}-a^g+1)r_1a^{n-g},$
then (\ref{eq9}) takes the form
$$
 Q_{n-1}\left(r_0+a^nr_1\right)+Q_n\left(z-2r_0-a^{n-g}r_1-a^{n+g+1}r_1\right)+
$$
\begin{equation}
\label{eq24}
 Q_{n+2}\left(2r_0+a^{n+1-g}r_1+a^{n+g+2}r_1-z\right)-Q_{n+3}\left(r_0+a^{n+2}r_1\right)=0.
\end{equation}
Let
$Q_n=B_ga^{gn}+B_{g-1}a^{(g-1)n}+\ldots+B_1a^n+B_0,$ $B_i=B_i(z).$
We introduce the notation $y=a^n$, then $Q_n=B_gy^g+\dots+B_0$, and (\ref{eq24}) takes the form
$$
 \sum_{s=0}^gB_s(a^{-g-s}(a^g-a^s)(a^{g+s+1}-1)(a^{2s+1}-1)r_1y^{s+1}-
a^{-s}(a^{2s}-1)((a^s-1)^2r_0+a^sz)y^s)=
$$
$$
 \sum_{s=1}^gy^s(B_sa^{-s}(1-a^{2s})((a^s-1)^2r_0+a^sz)+B_{s-1}a^{1-g-s}(a^g-a^{s-1})(a^{g+s}-1)(a^{2s-1}-1)r_1)=0.
$$
Hence we obtain
$$
 B_{s-1}=B_s\frac{a^{-s}(a^{2s}-1)((a^s-1)^2r_0+a^sz)}{a^{1-g-s}(a^g-a^{s-1})(a^{g+s}-1)(a^{2s-1}-1)r_1},\qquad s=1,\dots,g.
$$
Thus we found the polynomial $Q_n,$ satisfying (\ref{eq9}). Theorem 3 is proved.

${\ }$

\noindent The operator $L^{^\checkmark}_4$ is a discrete analogue of
$$
 {\cal L}^{^\checkmark}_4=(\partial_x^2+r_1a^x+r_0)^2+g(g+1)r_1a^x
$$
from \cite{Dav}, which commutes with a differential operator of order $4g+2$.

\subsection{Theorem 4}
Let $V_n=r_1\cos(n)+r_0,$ $W_n=-4r_1\sin(\frac{g}{2})\sin(\frac{g+1}{2})\cos(n+\frac{1}{2}).$
Equation (\ref{eq9}) takes the form
\begin{equation}
\label{eq25}
\begin{array}{c}Q_{n-1}\left(r_0+r_1\cos(n)\right)+
 Q_n\left(z-2r_0-2r_1\cos(g+\frac{1}{2})\cos(n+\frac{1}{2})\right)-\\
 Q_{n+2}\left(z-2r_0-2r_1\cos(g+\frac{1}{2})\cos(n+\frac{3}{2})\right)-
 Q_{n+3}\left(r_0+r_1\cos(n+2)\right)=0.\\
\end{array}
\end{equation}
Let us take the following ansatz
$$
 Q_n=A_g\cos(gn)+A_{g-1}\cos((g-1)n)+\ldots+A_1\cos(n)+A_0, \qquad A_i=A_i(z).
$$
We substitute $Q_n$ in (\ref{eq25}) and after some simplifications we obtain
$$
 A_0=A_1\frac{(z-2r_0+2r_0\cos(1))\sin(1)}{2r_1(\cos(g+\frac{1}{2})-\cos(\frac{1}{2}))\sin(\frac{1}{2})},
$$
$$
 A_{s-1}=\frac{A_s(z-2r_0+2r_0\cos(s))\sin(s)+A_{s+1}r_1(\cos(s-\frac{3}{2})-\cos(g+\frac{1}{2}))\sin(s-\frac{3}{2})}
 {r_1(\cos(g+\frac{1}{2})-\cos(s-\frac{1}{2}))\sin(s-\frac{1}{2})},
$$
where $2\leq s \leq g$, $A_{g+1}=0$, $A_g$ is a suitable constant. We found $Q_n$ satisfying (\ref{eq9}). Theorem 4 is proved.

The operator $L^{^\natural}_4$ is a discrete analogue of
$$
 {\cal L}^{^\natural}_4=(\partial_x^2+r_1\cos(x)+r_0)^2+g(g+1)r_1\cos(x)
$$
from \cite{Mir2}, which commutes with a differential operator of order $4g+2$.

Let us consider several examples.

\vspace{0.4cm}

\noindent{\bf Example 1} We introduce the notation $f(n)=r_3n^3+r_2n^2+r_1n+r_0$. The operator
$$
 L^{^\sharp}_4=(T+f(n)T^{-1})^2+2r_3n
$$
commutes with
$$
 L^{^\sharp}_6=T^3+3(f(n)+f'(n)+f''(n)+4r_3)T+3(f(n)+3r_3n+r_2)T^{-1}+
$$
$$
 (f(n-2)f'(n)+2f''(n)-8r_3)(f(n)-f'(n)+3r_3n-r_3+r_2)f(n)T^{-3}.
$$
The spectral curve is
$$
 w^2=z^3+(2r_2+3r_3)z^2+(r_1r_3+(r_2+r_3)(r_2+3r_3))z+
 r_3((r_2+r_3)(r_1+r_2+r_3)-r_0r_3).
$$

\vspace{0.4cm}

\noindent{\bf Example 2} The operator
$$
 L^{^\checkmark}_4=(T+(r_1a^n+r_0)T^{-1})^2+r_1(a^{3}-a^{2}-a+1)a^{n-1}
$$
commutes with
$$
 L^{^\checkmark}_6=T^3+\left(r_0(a+1+a^{-1})+r_1a^{n-1}(a^4+a^2+1)\right)T+(a+1+a^{-1})(r_1a^n+r_0)\times
$$
$$
 (r_1a^{n+1}-r_1a^n+r_1a^{n-1}+r_0)T^{-1}+
 (r_1a^n+r_0)(r_1a^{n-1}+r_0)(r_1a^{n-2}+r_0)T^{-3}.
$$
The spectral curve is
$
 w^2=z^3+\frac{2r_0(a-1)^2}{a}z^2+\frac{r_0^2(a-1)^4}{a^2}z.
$

\vspace{0.4cm}

\noindent{\bf Example 3} The operator
$$
 L^{^\natural}_4=(T+(r_1\cos(n)+r_0)T^{-1})^2-4r_1\sin(\frac{1}{2})\sin(1)\cos(n+\frac{1}{2})
$$
commutes with
$$
 L^{^\natural}_6=T^3+\left(2\cos(1)+1)(r_1(2\cos(1)-1)\cos(n+1)+r_0\right)T+
$$
$$
(2\cos(1)+1)(r_1\cos(n)+r_0)(r_1(2\cos(1)-1)\cos(n)+r_0)T^{-1}+
$$
$$
(r_1\cos(n-1)+r_0)(r_1\cos(n-2)+r_0)(r_1\cos(n)+r_0)T^{-3}.
$$
The spectral curve is
$
 w^2=z^3-8r_0\sin^2(\frac{1}{2})z^2-8(r_1^2(\cos(1)+1)-2r_0^2)\sin^4(\frac{1}{2})z.
$

${\ }$

\noindent{\bf Remark} We see that the pairs of commuting operators $L_4, L_{4g+2}$ and ${\cal L}_4, {\cal L}_{4g+2}$
have similar properties, and there are similar examples of such operators. It would be interesting to explain this duality.
So far, our attempts to do so by some discretization were not successful.

${\ }$

\noindent{\bf Acknowledgements} The authors are supported by a Grant of the Russian Federation for
the State Support of Researches (Agreement No 14.B25.31.0029).

\section*{Appendix}

Consider the differential--difference system (\ref{eq26}), (\ref{eq27}).

${\ }$

\noindent\textbf{Theorem 5} {\it
Let us assume that the potentials $V_n(t),\ W_n(t)$ of  $L_4=(T+V_n(t)T^{-1})^2+W_n(t)$ satisfy (\ref{eq26}), (\ref{eq27}).
We additionally assume that $[L_4,L_{4g+2}]=0$ for some difference operator $L_{4g+2}$. Then  $Q_n(t)$ associated with $L_4$
satisfies the evolution equation
\begin{eqnarray}
\label{eq29}
\dot{Q}_n=V_n(Q_{n+1}-Q_{n-1}).
\end{eqnarray}}
${\ }$

\noindent Equation (\ref{eq29}) defines the symmetry of (\ref{eq8}).
At $g=1$ functions $V_n(t),\ W_n(t)$ can be expressed via $\gamma_n(t)$ using (\ref{eq9}).
In this case the system (\ref{eq26}), (\ref{eq27}) and the equation (\ref{eq29}) are reduced to one equation
\begin{equation}
\label{eq30}
\dot{\gamma}_n=\frac{F_1(\gamma_n)(\gamma_{n-1}-\gamma_{n+1})}{(\gamma_{n-1}-\gamma_n)(\gamma_{n}-\gamma_{n+1})}.
\end{equation}
This equation is a discrete analogue of the Krichever--Novikov equation,
which appeared in the theory of rank two solutions of KP \cite{KN1}.
Equations similar to (\ref{eq26}), (\ref{eq27}) and (\ref{eq30}) were considered in \cite{ASh}, \cite{LWY}.

${\ }$

\noindent\textbf{Proof} Using
$
 (\partial_t-V_{n-1}(t)V_n(t)T^{-2})\psi_n=0,
$
and (\ref{eq4}) let us express
$\dot{\psi}_{n-1},$ $\dot{\psi}_n,$ $\dot{\psi}_{n+1},$ $\psi_{n-2},$ $\psi_{n-3}$ in terms of
$\psi_{n-1}, \psi_n, \chi_1(n), \chi_2(n)$
$$
\dot{\psi}_{n-1}=V_{n-2}V_{n-1}\psi_{n-3},\qquad
\dot{\psi}_n=V_{n-1}V_n\psi_{n-2}, \qquad
\dot{\psi}_{n+1}=V_nV_{n+1}\psi_{n-1},
$$
$$
 \psi_{n-2}=\frac{\psi_n-\psi_{n-1}\chi_2(n-1)}{\chi_1(n-1)},
$$
$$
 \psi_{n-3}=\frac{\psi_{n-1}(\chi_1(n-1)+\chi_2(n-2)\chi_1(n-1))-\psi_n\chi_2(n-2)}{\chi_1(n-2)\chi_1(n-1)}.
$$
From (\ref{eq4}) it follows that
$$
\dot{\psi}_{n+1}-\chi_1(n)\dot{\psi}_{n-1}-\dot{\chi}_1(n)\psi_{n-1}-\chi_2(n)\dot{\psi}_n-\dot{\chi}_2(n)\psi_n={\cal A}_n\psi_n+{\cal B}_n\psi_{n-1}=0,
$$
where
$$
 {\cal A}_n=V_{n-2}V_{n-1}\chi_1(n)\chi_2(n-2)-\chi_1(n-2)(V_{n-1}V_n\chi_2(n)+\chi_1(n-1)\dot{\chi}_2(n)),
$$
$$
 {\cal B}_n=V_{n-2}V_{n-1}\chi_1(n)\left(\chi_1(n-1)+\chi_2(n-2)\chi_2(n-1)\right)+
$$
$$
+V_n\chi_1(n-2)\left(V_{n+1}\chi_1(n-1)+V_{n-1}\chi_2(n-1)\chi_2(n)\right)-\chi_1(n-2)\chi_1(n-1)\dot{\chi}_1(n).
$$
Consequently, we have ${\cal A}_n={\cal B}_n=0$.
For $\chi_1=-V_n(t)\frac{Q_{n+1}(t)}{Q_n(t)}$, $\chi_2=\frac{w}{Q_n(t)}$ it follows from
(\ref{eq8}) and from ${\cal A}_n={\cal B}_n=0$ that $Q_n(t)$ satisfies (\ref{eq29}). Theorem 5 is proved.

\vspace{0.4cm}

\noindent {\it G.S. Mauleshova}, Novosibirsk State University, Russia

\noindent e-mail: guna$_{-}$1986@mail.ru

\vspace{0.4cm}

\noindent {\it A.E. Mironov}, Sobolev Institute of Mathematics, Novosibirsk, Russia and

\noindent Laboratory of Geometric Methods in Mathematical Physics, Moscow State University

\noindent e-mail: mironov@math.nsc.ru

\end{document}